\documentclass[aip,jcp,
 amsmath,amssymb,
reprint, 
]{revtex4-2}
\usepackage[utf8]{inputenc}
\usepackage[version=4]{mhchem} 
\usepackage{bm}
\usepackage{amsmath}
\usepackage{MnSymbol}
\usepackage{physics}
\usepackage{braket}
\usepackage{xcolor}
\usepackage{graphicx} 

\begin{document}
\title[LR-CCSD-PBC]{Frequency Dependent Polarizability for 1D Periodic Systems at Coupled Cluster with Single and Double Excitations Level}
\author{Marco Caricato}
\email{mcaricato@ku.edu}
\affiliation{Department of Chemistry, University of Kansas, 1567 Irving Hill Road, Lawrence, Kansas 66045, United
  States}

\author{Taylor Parsons}%
 \affiliation{Department of Chemistry, University of Kansas, 1567 Irving Hill Road, Lawrence, Kansas 66045, United
  States}%

\author{Michael J. Frisch}%
 \affiliation{Gaussian, Inc., 340 Quinnipiac St Bldg 40
Wallingford, CT 06492 USA}%

\author{Julia Abdoullaeva}
 \affiliation{Department of Chemistry, University of Kansas, 1567 Irving Hill Road, Lawrence, Kansas 66045, United
  States}

\date{\today}

\begin{abstract}
We report the first implementation of the frequency-dependent electric dipole-electric dipole polarizability for 1D periodic systems computed with the coupled cluster with single and double excitations (CCSD) method with periodic boundary conditions (PBCs). The implementation is performed in the CCResPy open-source software, based on Python and the NumPy library. The complete equations and many details of the implementation are discussed. The test calculations show the impact on this linear response property of passing from a single molecule to a periodic chain, where the relative magnitude of the polarizability tensor elements is inverted. This work also explores the convergence towards the PBC thermodynamic limit with $k$-space sampling, and some remaining issues in the definition of the electric dipole operator for periodic systems. This work represents a significant step forward for the simulation of optical response properties for solid-state materials with accurate and systematically improvable quantum mechanical methods.
\end{abstract}

\maketitle

\section{Introduction\label{into}}

The simulation of the electronic response of solid state materials to external fields is paramount for the interpretation and prediction of electronic and optical properties  of these systems and their technological applications. The theoretical workhorse for these simulations is density functional theory (DFT) with periodic boundary conditions (PBC), which provides the best balance between reliability and computational cost.\cite{lcaobook} However, 
DFT for the solid state suffers from the same issues as for molecules: the exact functional is unknown and any given approximate functional is not systematically improvable. Furthermore, a PBC simulation is inherently more computationally demanding than that on a molecular system of the same size as the simulation cell, which further limits the quality of the functional and basis set expansion that can be used compared to molecules. Benchmarking of approximate DFT methods is harder for solids than for molecules, because a direct comparison with experimental data is difficult, and high-level simulations (which are often used as reference in molecular benchmarking studies) are not available. 

For isolated molecules, when highly-accurate data are needed, the wave function (WF) formalism is invoked because systematically improvable electron correlation methods built on top of a reference WF can be formulated. Except for near-degeneracy situations where multi-reference methods are required, the reference WF is provided by the Hartree-Fock (HF) method. From there, post-HF methods lead towards the exact WF (within a set of constraints: the Born-Oppenheimer approximation, a finite and often incomplete basis set expansion, and a non-relativistic electronic Hamiltonian) with a sequence of well-defined improvements of increasing complexity and computational effort. The HF wave function is the best approximation, in a variational sense, of the exact WF when using a single Slater determinant built from optimized molecular orbitals (MOs). Post-HF methods improve upon this description using a linear or non-linear combination of Slater determinants built as a set of single, double, etc. electron excitations from occupied to unoccupied MOs. The exact WF is obtained when all possible excitations are accounted for within a finite basis set. Although the exact solution is practically unachievable for most real cases, a systematic path towards more accurate results obtained through the inclusion of more and more excited determinants is obviously a very desirable characteristic of these methods. Among these post-HF methods, those based on coupled cluster (CC) theory are the most successful in quantum chemistry.\cite{Bartlett:2007bh, MBPT} CC theory uses a non-linear expansion of the WF based on an exponential excitation operator $e^{\hat T}$, and approximate CC methods are defined in terms of the classes of excitations that are included in the $\hat T$ cluster operator. The most common CC method includes single and double excitations (CCSD), because it provides the first important contribution to the correlation energy while maintaining a reasonable computational cost: $O(N^6)$ where $N$ is a measure of the system size. CC's popularity stems from including higher-order excitations for a specific level of truncation in the residual equations (e.g., CCSD includes up to quadruple excitations thanks to products of single and double excitation operators) and the energy is size-extensive at each level of truncation, thus ensuring a balanced increase of the error with system size.\cite{Bartlett:2007bh, MBPT} The downsides of CC theory are that the complexity of the WF expansion grows quickly, both in terms of equations and of the corresponding computer code, and that the computational scaling is unfavorable compared to DFT: $O(N^6)$ for CCSD vs $O(N^3)$ for standard density functional implementations. 

Despite the steep computational cost, CC methods have been extended to the solid state, from the seminal work of Hirata \textit{et al.}\cite{hirata_coupled-cluster_2004,ohnishi_hybrid_2011,
  gilliard_second-order_2014, keceli_fast_2010,
  sode_coupled-cluster_2009, hirata_thermodynamic_2012,
  hirata_many-body_2000} on CCSD energies to the more recent efforts from the Chan, Gr\"uneis, Berkelback, and other groups.\cite{white_coupled_2020,
  pulkin_first-principles_2020, peng_conservation_2021,
  white_finite-temperature_2020, gao_electronic_2020,
  white_time-dependent_2019, zhu_coupled-cluster_2019,
  white_time-dependent_2018, mcclain_gaussian-based_2017,
  mcclain_spectral_2016,gallo_periodic_2021, irmler_particle-particle_2019,  zhang_coupled_2019, gruber_applying_2018, gruneis_coupled_2018,  gruber_ab_2018, hummel_low_2017, liao_communication_2016,  gruneis_coupled_2015, gruneis_natural_2011, shepherd_many-body_2013,
  schafer_local_2021, linlin:2024, XING2024112755} These developments include the calculation of ground and excited state energies,\cite{pulkin_first-principles_2020,
  mcclain_gaussian-based_2017, Kosugi:2019, lange_active_2020,gallo_periodic_2021, wang_excitons_2020,  katagiri_equation--motion_2005, Laughon:2022, wang_absorption_2021, Ye:2024, Neufeld:2023, Neufeld:2022, Vo:2024} and provide much needed high-level reference data for more approximate methods. These works also provide an opportunity to directly explore the importance of electron correlation in electronic properties of materials, going beyond the independent-particle approximation often invoked in condensed-phase physics. 

Excitation energies are important to predict the optical band gap in solid materials. However, the motion of the electron density in response to an external field is more generally described by the full linear response (LR) function.\cite{Christiansen:1998wz, helgaker_recent_2012} In this work, we present the first implementation for 1D periodic systems at CCSD level of the frequency-dependent LR function for the electric dipole operator in the length gauge (LG), i.e. the electric dipole-electric dipole polarizability tensor $\boldsymbol{\alpha}(\omega)$, where $\omega$ is the frequency of the external field. The implementation was carried out in the \texttt{CCResPy} open source software, based on the Python language and the NumPy module library.\cite{github-ccrespy} We report the equations and many details of the implementation, together with a set of proof-of-concept calculations on 1D chains of simple molecules. We explore how these electronic properties change going from an isolated molecule to a periodic system, and we investigate the convergence of the calculations towards the thermodynamic limit. This work represents another fundamental step in the direction of performing simulations of electronic properties of condensed-phase systems with methods that are as accurate as those for molecules. 

This paper is organized as follows. Section \ref{theory} reports detailed equations and discusses various aspects of the implementation. Simulations on test systems are reported in section \ref{results}. A discussion of these data and concluding remarks are reported in section \ref{discuss}.

\section{Theory \label{theory}}

This section reports the theory for the evaluation of the frequency-dependent linear response function at CCSD level with 1D periodicity and implementation details in the \texttt{CCResPy} software.\cite{github-ccrespy} We start with a short review of response theory for the linear response of the electronic WF to an external field; a detailed account of response theory can be found in Ref. \citenum{Christiansen:1998wz}. The time evolution of the system is given by the time-dependent Schr\"odinger equation:
\begin{equation}
  \hat H | \Psi \rangle = i \partial/\partial t | \Psi \rangle
\end{equation}
where $\hat H = \hat H_0 + \hat V(t)$ is the electronic Hamiltonian, written as the sum of an unperturbed term and a time-dependent perturbation, oscillating with frequency $\omega$:
\begin{equation}
  \hat V(t) = \sum_y \epsilon_y (\omega) \hat Y e^{-i \omega t}
  \label{eq:pert}
\end{equation}
In Eq. \ref{eq:pert}, $\epsilon_y (\omega)$ is the strength of the field in the Cartesian direction $y$ and $\hat Y$ is a perturbation operator, e.g., the $y$-component of the electric dipole. The time-dependent expectation value for an observable $X$ can be expanded in orders of the perturbation $\hat V(t)$; for instance, up to first order:
\begin{equation}
  \langle \Psi | \hat X | \Psi \rangle = \langle \Psi_0 | \hat X | \Psi_0 \rangle + \sum_y \epsilon_y (\omega) \llangle X,Y \rrangle_\omega e^{-i \omega t} + O(2)
  \label{eq:expctX}
\end{equation}
where $\Psi_0$ is the unperturbed wave function, and the expansion coefficient $\llangle X,Y \rrangle_\omega$ is the linear response function. For $X$ and $Y$ equal to the electric dipole $\mu$, the LR function corresponds to the electric dipole-electric dipole polarizability tensor $\boldsymbol{\alpha}(\omega)$:
\begin{equation}
  \alpha_{\alpha\beta}(\omega) = \llangle \mu_\alpha,\mu_\beta \rrangle_\omega 
  \label{eq:alpha}
\end{equation}
Although the LR function $\llangle X,Y \rrangle_\omega$ can be written as a sum-over-states (SOS) formula over the eigenvalues and eigenvectors of $\hat H_0$, that is not a computationally efficient approach because the SOS series is slowly convergent, and the evaluation of the excited states of $\hat H_0$ is computationally expensive. Instead, a time-averaged quasi-energy $Q(t)$ can be defined, and the response functions at each order are computed as derivatives of $Q(t)$ with respect to the field strength parameters.\cite{Christiansen:1998wz} This approach requires the evaluation of perturbed amplitudes for the wave function expansion. In LR-CC theory, the response of the reference MO coefficients is usually neglected to avoid unphysical poles of the LR function, so one needs to evaluate the response of the amplitudes of the $\hat T$ excitation operator.\cite{KOCH:1990dq, Christiansen:1998wz, Christiansen:1998qe} The explicit expressions to evaluate the CCSD LR function are discussed in section \ref{sec:CCSD}.

\subsection{Periodic Boundary Conditions \label{sec:PBC}}

\begin{figure}[htp!]
    \includegraphics[width=8cm]{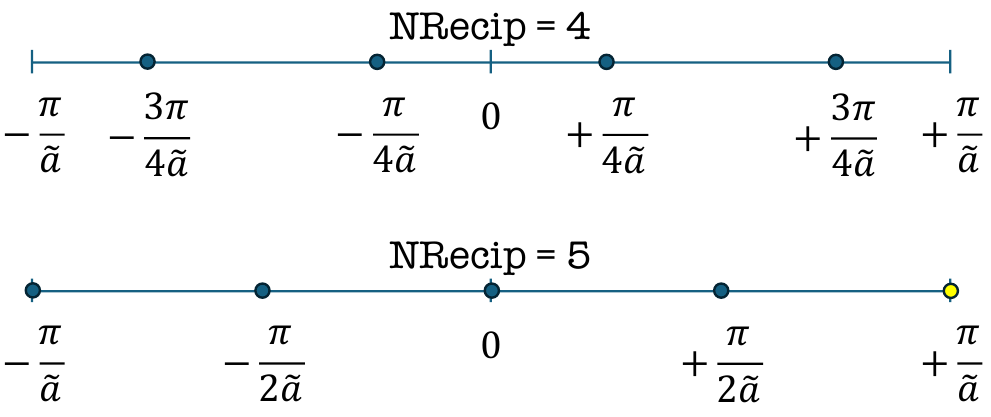}
    \caption{Graphical representation of the grid mesh for \texttt{NRecip=4} and \texttt{NRecip=5} in the first Brillouin zone for 1D periodicity. The blue circles represent the $k$ point selection; the yellow circle at $k=+\pi/\tilde a$ is not selected as it is equivalent to the other edge point at $k=-\pi/\tilde a$ for \texttt{NRecip=5}. \label{fig:nrecip}}
  \end{figure}

The \texttt{CCResPy} implementation is based on a basis of Gaussian-type atomic orbitals (GTOs), similarly to the seminal work of Hirata \textit{et al.}\cite{hirata_coupled-cluster_2004} on CCSD energy and of Scuseria \textit{et al.}\cite{Kudin:2000jr,Kudin1998611,Kudin:2000bq,Izmaylov:2006jb} for HF and DFT. \texttt{CCResPy} only solves the CCSD equations, while the data from the reference wave function are generated with the GAUSSIAN suite of programs.\cite{gdv} Specifically, we run a GAUSSIAN job at the HF level for the system of interest, and extract the relevant data via the \texttt{gauopen} functionality, an open-source program that extracts content from GAUSSIAN binary files and converts them to other formats (unformatted text or Python NumPy arrays in our case). The data needed by our program include: the crystal orbital coefficients in $k$ space (COk), the two-electron repulsion integrals (2ERIs) in real-space atomic orbital basis (AOr), the AOr Fock and overlap matrices, the AOr dipole integrals, the number of cells in real space ($N_C$) and the number of $k$ points ($N_k$). Because GAUSSIAN uses a symmetric $k$-point grid, we only need to know $N_k$ to generate the same grid in the first Brillouin zone defined within the $[-\pi/\tilde a,\pi/\tilde a)$ interval, where $\tilde a$ is the translation vector length. More specifically, we read the value of the keyword \texttt{NRecip=N}, which simultaneously indicates whether the edge and $\Gamma$ points of the Brillouin zone are included or not, depending on whether $N$ is odd or even, respectively, and the total number of $k$ points in the mesh. In the odd $N$ case, the mesh has one less point because the $-\pi/\tilde a$ and $+\pi/\tilde a$ edge points are equivalent, resulting in an effective $N_k=N-1$ number of $k$ points. For example, \texttt{NRecip=4} and \texttt{NRecip=5} correspond to different meshes but they both have $N_k=4$, as shown in Figure \ref{fig:nrecip}. Other choices of grid mesh could be easily implemented in the future.

The equations are implemented in terms of spin-orbitals, but we do not perform spin integration beforehand: the tensor elements corresponding to forbidden spin blocks are simply set to zero. This is not efficient, but it significantly simplifies the implementation. Although the simulations in this work are limited to closed-shell cases with canonical orbitals, the code can in principle handle open-shell systems and non-canonical orbitals with a few modifications.

In the current implementation, the CCSD equations are solved completely in $k$ space, so that all AOr matrices are first converted to COk basis as discussed in Refs. \citenum{hirata_coupled-cluster_2004,Izmaylov:2006jb}. Complex numbers are represented with 128-bit precision and real numbers with 64-bit precision. Here we only note that the single and double excitation amplitudes as well as the 2ERIs in COk basis need to respect the momentum conservation conditions:\cite{hirata_coupled-cluster_2004}
\begin{equation}
\begin{split}
& (k_a - k_i)\tilde a = 2\pi m_1 \\
& (k_a + k_b - k_i - k_j)\tilde a = 2\pi m_2
\end{split}
\label{eq:momentum}
\end{equation}
where $k_p$ represents the $k$ point for the $p$-th CO, $i,j$ represent occupied COs, $a,b$ unoccupied COs, and $m_1,m_2$ are two integers. 
The relations in Eq. \ref{eq:momentum} indicate that the choice of $k$ points across excitations is not completely arbitrary: there are only $N_k$ independent $k$ points for one-electron (1E) quantities and $N_k^3$ for two-electron (2E) quantities. However, in the current implementation we still treat 1E and 2E tensors as $N^2_k$ and $N^4_k$ quantities, respectively, even if many of their elements are zero because of momentum conservation. This is obviously inefficient in terms of storage and computational scaling, but it has the distinct advantage that it allows us to treat the tensors in the molecular and periodic cases on the same footing by defining a collective index for the orbital and its $k$ point: $P\equiv pk_p$. Examples for the CCSD single and double excitation $t$ amplitudes are shown in Table \ref{tab:index}.
\begin{table}[h!]
    \centering
    \begin{tabular}{cc}
    \hline
       Periodic  & Molecular \\
       \hline
 $t_{ik_i}^{ak_a} \equiv t_I^A$ &$t_i^a$ \\
 $t_{ik_ijk_j}^{ak_abk_b} \equiv t_{IJ}^{AB}$ &$t_{ij}^{ab}$ \\
 \hline
    \end{tabular}
    \caption{Labeling of single and double excitation amplitudes for the periodic and the molecular cases.}
    \label{tab:index}
\end{table}

Because our implementation is based on Python and the tensor contractions are performed with the NumPy \texttt{einsum} function, which automatically handles real and complex algebra, our CCSD code is essentially transparent to whether the calculation is periodic or molecular. The information on the type of calculation is limited to the part of the code that performs the AOr-to-COk transformation of the 1E matrices and 2ERIs. Once these COk quantities are created with the appropriate zero values due to the momentum conservation constraints in Eq. \ref{eq:momentum}, all the CCSD tensors will automatically respect these constraints and no explicit summation over $k$ points is required. This makes the implementation of the CCSD part of the code considerably cleaner, easier to further develop to add new features, and with a less steep learning curve for new developers.

\subsection{LR-CCSD Equations \label{sec:CCSD}}

In this section, we report the complete equations for the LR-CCSD method for the evaluation of the electric dipole-electric dipole polarizability tensor. As discussed in section \ref{sec:PBC}, these equations apply equally to the periodic and molecular cases because of the use of collective indices. Although the equations for molecular LR-CCSD are known,\cite{KOCH:1990dq, Christiansen:1998wz, Christiansen:1998qe} we report them here explicitly  exactly how they appear in \texttt{CCResPy}\cite{github-ccrespy}, which allows us to discuss details of the implementation. The choice of intermediates is based on the work of Gauss \textit{et al.}\cite{GAUSS:1991wd} and we used their notation conventions. However, there are a few notable differences in some choices of contractions (for computational expediency) and in the order of the tensors' indices with respect to Ref. \citenum{GAUSS:1991wd}. The latter is due to the fact that  the CCSD-PBC equations are solved in $k$ space where the tensors are complex, therefore, some of the integral and tensor symmetries are lost compared to purely real quantities and following the correct order of the indices is essential.

In the following, a summation over common indices is implied, and parentheses are used to indicate the priority of contraction. The 2ERIs are reported in the physicist notation, and the double bar indicates anti-symmetrization: $\langle PQ||RS \rangle = \langle PQ|RS \rangle - \langle PQ|SR \rangle$. We used the common convention that the $\{I,J,\dots\}$ collective indices refer to occupied orbitals and $\{A,B,\dots\}$ to unoccupied orbitals. The current implementation maintains all arrays in memory, except for the $N_V^4$ arrays ($N_V=N_vN_k$, where $N_v$ is the number of unoccupied orbitals), which are stored on binary files that are loaded when needed for their contraction. The order of the indices in the NumPy arrays follows the order used here in the text of the equations, e.g., $\langle IJ||AB\rangle \rightarrow \texttt{IJAB[I,J,A,B]}$ for integrals, $t_{IJ}^{AB} \rightarrow \texttt{t2[I,J,A,B]}$ for the amplitudes, and similarly for other 1E and 2E integral and excitation amplitude quantities. For de-excitation amplitudes ($\lambda$ and $\xi$), the array ordering is the same as above but the equation notation is opposite to that of the excitation amplitudes, e.g., $\lambda_{AB}^{IJ} \rightarrow \texttt{l2[I,J,A,B]}$, for consistency with the literature.\cite{GAUSS:1991wd} Since the \texttt{einsum} function allows us to write the contractions virtually as written on paper, the explicit expressions reported below are essentially exactly how they are implemented in the code.

The equations for the correlation energy and for the $t$ amplitudes are:

\begin{equation}
\label{eq:ene}
E_{corr} = \frac{1}{N_k} t_I^{A} f_{IA} + \frac{1}{4N_k^3} \tau_{IJ}^{AB} \langle IJ||AB \rangle
\end{equation}

\begin{widetext}
\begin{equation}
\label{eq:t1}
t_I^{A}{D_I^{A}} =  
f_{IA} + t_I^{E} {\cal F}_{AE} - t_M^{A} {\cal F}_{MI} 
+ \frac{1}{N_{k}} t_{IM}^{AE} {\cal F}_{ME} 
- \frac{1}{2N_{k}^2} t_{IM}^{EF} \langle MA||EF \rangle 
+ \frac{1}{2N_{k}^2} t_{MN}^{AE} \langle NM||IE \rangle 
 + \frac{1}{N_{k}} t_N^{F} \langle NA||FI \rangle 
\end{equation}

\begin{equation}
\label{eq:t2}
\begin{split}
t_{IJ}^{AB} D_{IJ}^{AB} = &  
\langle IJ||AB \rangle^* + P_{AB} \left [ \left ( {\cal F}_{BE} - \frac{1}{2}t_M^B{\cal F}_{ME} \right ) t_{IJ}^{AE}  
- t_M^{A} \langle IJ||MB \rangle^* \right ]\\
&- P_{IJ} \left [ \left ( {\cal F}_{MJ} + \frac{1}{2}t_J^E{\cal F}_{ME} \right ) t_{IM}^{AB} 
+ t_I^{E} \langle JE||AB \rangle^*  \right ]
+ P_{IJ}P_{AB} \left [ \frac{1}{N_k} t_{IM}^{AE} W_{MBEJ} 
- t_M^{A} \left ( t_I^{E} \langle MB||EJ \rangle \right ) \right ]\\
&+ \frac{1}{2N_{k}}  \tau_{IJ}^{EF} \langle AB|| EF\rangle 
+ \frac{1}{2N_{k}}  \tau_{MN}^{AB} W_{MNIJ} 
- P_{AB} \left [ t_M^{A} \left ( \frac{1}{2N_k} \tau_{IJ}^{EF} \langle MB||EF \rangle \right ) \right ]
\end{split}
\end{equation}
\end{widetext}
where $P_{PQ}$ is an anti-symmetrization operator: $P_{PQ} X_{PQ}= X_{PQ} -X_{QP}$ and the orbital energy ($\varepsilon_P$) difference terms are: 
\begin{equation}
\label{eq:D}
\begin{split}
& D_I^A = \varepsilon_I- \varepsilon_A \\
& D_{IJ}^{AB} = \varepsilon_I + \varepsilon_J - \varepsilon_A - \varepsilon_B\\
\end{split}
\end{equation}

The intermediates used in Eqs. \ref{eq:ene}-\ref{eq:t2} are reported in Eqs. S1-S6 of the Supporting Information (SI). Contrary to Ref. \citenum{GAUSS:1991wd}, we do not define a $W_{ABEF}$ intermediate because it would be an extra $N_V^4$ array to store. Instead, we perform the $\langle AB||EF \rangle$ contraction directly and add the last term in Eq. \ref{eq:t2}, which scales as $N_O^3N_V^3$ ($N_O=N_oN_k$, where $N_o$ is the number of occupied COs). The last term in the intermediate in Eq. S4 also has an extra factor of 2 compared to the equivalent term in Ref. \citenum{GAUSS:1991wd}, to account for the last term of the missing $W_{ABEF}$ intermediate.

The $\lambda$ equations are:

\begin{widetext}
\begin{equation}
\begin{split}
D_I^A \lambda_A^I = &  F_{IA} + \lambda_E^I \tilde{\cal F}_{EA} - \lambda_A^M \tilde{\cal F}_{IM} + \frac{1}{N_k}\lambda_E^M \tilde W_{IEAM} 
  + \frac{1}{2N_k^2} \lambda^{IM}_{EF} \tilde W_{EFAM} - \frac{1}{2N^2_k} \lambda^{MN}_{AE}  \tilde W_{IEMN} \\
& + \frac{1}{N_k}G_{EF} \langle IE || FA \rangle 
+ \frac{1}{N_k} G_{MN} \langle IM || NA \rangle 
 + \frac{1}{N_k} \left ( t_M^F G_{FE} - t_N^E G_{MN} \right ) \langle IM || AE \rangle
\end{split}
\label{eq:l1}
\end{equation}

\begin{equation}
\begin{split}
D_{IJ}^{AB}\lambda^{IJ}_{AB} = & \langle IJ || AB \rangle + \frac{1}{2N_k} \lambda^{IJ}_{EF} \tilde W_{EFAB} + \frac{1}{2 N_k} \lambda^{MN}_{AB} \tilde W_{IJMN}  + P_{IJ}P_{AB} \left[ \frac{1}{N_k}\lambda^{IM}_{AE} \tilde W_{JEBM} + N_k\lambda_A^I F_{JB} \right] \\
& + P_{AB} \left[ \left ( G_{BE} - \lambda _B^M t_M^E \right ) \langle IJ || AE \rangle - \lambda_A^M \langle IJ || MB \rangle + \lambda^{IJ}_{AE} \tilde F_{EB} \right] \\
& + P_{IJ} \left[ \left ( G_{MJ} + \lambda _E^J t_M^E \right )  \langle IM || AB \rangle + \lambda_E^I \langle JE || AB \rangle + \lambda^{IM}_{AB} \tilde F_{JM} \right] 
\end{split}
\label{eq:l2}
\end{equation}
\end{widetext}

The new intermediates are reported in Eqs. S7-S14 of the SI. $\tilde W_{IJMN}$ is identical to $W_{IJMN}$ in Eq. S4 because we defined the latter differently than in Ref. \citenum{GAUSS:1991wd} to avoid the need for a $W_{ABEF}$ intermediate, as explained above; nonetheless, we preserve the \textit{tilde} notation for consistency. Contrary to the $t$ amplitudes case, it makes sense to build a $N_V^4$ intermediate: $\tilde W_{ABEF}$  (in Eq. S9 of the SI) as this is only computed once (after the $t$ equations are solved) and stored on disk. First, the $\langle AB||EF \rangle$ binary file is moved to a $\tilde W_{ABEF}$ binary file; the latter is then loaded in memory and the other terms in Eq. S9 are added to it; finally, the array is stored back on the $\tilde W_{ABEF}$ binary file and the memory array is released after its contribution to the $\tilde W_{EFAM}$ intermediate in Eq. S9 is evaluated. During the solution of the $\lambda _{AB}^{IJ}$ (and $\bar t _{AB}^{IJ}$) equations, at each iteration the $\tilde W_{ABEF}$ intermediate is loaded in memory from disk, contracted as in Eq. \ref{eq:l2}, then the array is immediately released to reduce its impact on memory usage. Additionally, it is worth noting that it is faster to perform the contraction of the $t_1$ amplitudes twice explicitly in Eq. S9 (including the transposition: $\texttt{IABC[M,A,E,F]} \rightarrow \texttt{X[A,M,E,F]}$ to set the indices of the integral array in the correct order before the contraction), rather than performing one contraction plus a $P_{AB}$ operation on a $N_V^4$ array. Similarly, the $\tau_{MN}^{AB}$ and $\langle MN||EF \rangle$ terms (stored as \texttt{tau[M,N,A,B]} and \texttt{IJAB[M,N,E,F]} arrays, respectively) are permuted to move the $MN$ indices to the right-hand side \textit{before} the contraction, so that they are already in the right order (i.e.,  contraction indices as fast indices) before calling the \texttt{einsum} function. All of the \textit{tilde} intermediates are fixed and evaluated only once after the solution of the $t$ equations, Eqs. \ref{eq:t1}-\ref{eq:t2}.  These intermediates are also utilized in the linear response equations for the calculations of the perturbed $\bar t(X,\pm \omega)$ amplitudes as shown below. On the other hand, the two $G$ intermediates in Eqs. \ref{eq:l1}-\ref{eq:l2} need to be evaluated at each iteration because they include the $\lambda$ amplitudes (see Eq. S14 of the SI).

The evaluation of the linear response function requires the calculation of perturbed single and double excitation amplitudes, $\bar t_I^A(X,\pm \omega)$ and $\bar t_{IJ}^{AB}(X, \pm\omega)$ where $X$ is a particular perturbation (e.g., a Cartesian component of the electric dipole) and $\omega$ is the frequency of the external field. Two sets of amplitudes are needed for one frequency, where the sign of $\omega$ is related to the complex conjugation of the field strength amplitude.\cite{Christiansen:1998wz} In the following equations, we omit $(X,\pm\omega)$ for clarity. These amplitudes are computed by solving linear systems of equations where the right-hand side is represented by similarity-transformed perturbation integrals in COk basis:

\begin{equation}
\begin{split}
R_I^A(X) =& X_{IA} + \frac{1}{N_k}t_{IK}^{AC} X^*_{KC}  - t_I^C t_K^A X^*_{KC}  \\
&  + t_I^C X_{CA} - t_K^A X_{IK} \\
R_{IJ}^{AB}(X) = & -P_{IJ} \left[ \left( X_{IK} + t_I^C X_{KC}^* \right) t_{KJ}^{AB} \right] \\
&+ P_{AB} \left[ \left( X_{CB} - t_K^B X^*_{KC} \right) t_{IJ}^{AC} \right]
\end{split}
\label{eq:rhs}
\end{equation}
where $X_{PQ}$ are the perturbation integrals in COk basis. For the electric dipole operator in the length gauge, these integrals are obtained as described in Refs. \citenum{Champagne:1992fj, Kudin:2000bq, Balduf:2022, Rerat:2021}. Rearranging the $\bar t(X,\pm\omega)$ perturbed amplitude equations in a form that is compatible with a standard iterative solution (as for the $\lambda$ amplitudes in Eqs. \ref{eq:l1}-\ref{eq:l2}), they become:

\begin{widetext}
\begin{equation}
\begin{split}
\bar{t}_I^A(D_I^A \mp \omega) = & -R_I^A+ \bar{t}_I^E \tilde {\cal F}_{AE} - \bar{t}_M^A \tilde {\cal F}_{MI}  + \frac{1}{N_k}\bar{t}_M^E \tilde W_{MAEI} 
+ \frac{1}{N_k}\bar{t}_{IM}^{AE} {\cal F}_{ME} + t_I^B \tilde G_{AB} - t_J^A \tilde G_{JI} \\
& - \frac{1}{2N^2_k} \bar{t}_{IM}^{EF} \langle MA || EF \rangle 
+ \frac{1}{2N^2_k} \bar{t}_{NM}^{EA} \langle NM||IE \rangle 
\end{split}
\label{eq:tx1}
\end{equation}

\begin{equation}
\begin{split}
\bar{t}_{IJ}^{AB} (D_{IJ}^{AB} \mp \omega) = & -R_{IJ}^{AB} +  \frac{1}{2N_k} \bar{t}_{IJ}^{EF} \tilde W_{ABEF} + \frac{1}{2N_k} \bar{t}_{MN}^{AB} \tilde W_{MNIJ}  + P_{IJ}P_{AB} \left[ \frac{1}{N_k}\bar{t}_{IM}^{AE} \tilde W_{MBEJ} \right] \\
& + P_{IJ} \left [ \bar{t}_I^C \tilde W_{ABCJ} - \bar{t}_{IM}^{AB} \tilde {\cal F}_{MJ}  
  -\left ( \tilde G_{MJ} + t_J^D Y_{MD} - \frac{1}{N_k}\bar t _K^C  \langle KM||JC \rangle \right ) t_{IM}^{AB} \right ] \\
& + P_{AB}  \left [ \bar{t}_{IJ}^{AE} \tilde {\cal F}_{BE} - \bar{t}_K^A \tilde W_{KBIJ} 
  + \left ( \tilde G_{BD} - t_M^B Y_{MD} + \frac{1}{N_k} \bar t_K^C \langle KB||CD \rangle \right ) t_{IJ}^{AE} \right ] \\
\end{split}
\label{eq:tx2}
\end{equation}
\end{widetext}
where the $\tilde {\cal F}$ and $\tilde W$ intermediates are defined in Eqs. S7-S12 of the SI and the new intermediates are reported in Eq. S15 of the SI. For a vector perturbation like the electric dipole operator, six sets of single and double excitation amplitudes are computed (one for each Cartesian component of the dipole times two for the $\pm \omega$ frequencies). However, we implemented a check on the magnitude of the dipole integrals, so that no amplitudes are computed if a component of the dipole is zero in a particular direction.

Once the perturbed amplitudes are available, a new set of constant terms is evaluated: $\xi_A^I$ and $\xi_{AB}^{IJ}$, which are similar to those needed in the EOM-CCSD gradients.\cite{STANTON:1993nx, STANTON:1994eu, Caricato:2013jg, Caricato:2012ida} The explicit equations for the $\xi_A^I$ and $\xi_{AB}^{IJ}$ tensors and the corresponding intermediates are reported in Eqs. S16-S21 of the SI. The $\xi_A^I$ and $\xi_{AB}^{IJ}$ tensors in Eqs. S16-S17 are evaluated simultaneously so that the arrays for the common intermediates in Eqs. S18-S21 are computed and used once, then immediately discarded.
The last piece that is needed for the evaluation of the frequency-dependent response function is a one-particle transition density-like (1PDM-like) tensor, $\rho_{PQ}(X,\pm \omega)$, reported in Eq. S22 of the SI.

The linear response function is finally obtained as:\cite{KOCH:1990dq, Christiansen:1998wz, helgaker_recent_2012}
\begin{widetext}
\begin{equation}
\begin{split}
\llangle  X,Y\rrangle _\omega = & -\frac{1}{N_k} \xi_A^I(X,+\omega) \bar t_I^A(Y,-\omega) - \frac{1}{4N^3_k}\xi_{AB}^{IJ}(X,+\omega) \bar t_{IJ}^{AB}(Y,-\omega) \\
& +\frac{1}{N_k} \left [ X^*_{IJ}\rho_{IJ}(Y,+\omega) + X^*_{IA}\rho_{IA}(Y,+\omega) +X^*_{AB}\rho_{AB}(Y,+\omega) \right ] \\
& +\frac{1}{N_k} \left [ X^*_{IJ}\rho_{IJ}(Y,-\omega) + X^*_{IA}\rho_{IA}(Y,-\omega) +X^*_{AB}\rho_{AB}(Y,-\omega) \right ] \\
\end{split}
\label{eq:lrf}
\end{equation}
\end{widetext}
This is the symmetric formulation of the LR function, which requires only the evaluation of the $\bar t(X,\pm\omega)$ amplitudes. An equivalent formulation can be obtained in a non-symmetric form that also requires the evaluation of perturbed $\lambda$ amplitudes $\bar \lambda(X,\pm\omega)$,\cite{Stanton:2000} but we have not implemented it yet. In this work, the elements $X,Y$ represent the Cartesian components of the electric dipole operator in the length gauge.

\subsection{Further Implementation Details  \label{sec:code}}

The amplitudes equations, Eqs. \ref{eq:t1}, \ref{eq:t2}, \ref{eq:l1}, \ref{eq:l2}, \ref{eq:tx1}, and \ref{eq:tx2}, are solved iteratively by adjourning the amplitudes on the left-hand side with the residuals on the right-hand side (divided by the orbital energy denominators in Eq. \ref{eq:D}), computed with the amplitude from the previous iteration. Convergence is achieved when a set of criteria is satisfied: $|E^{[n]}-E^{[n-1]}|< 10^{-8}$ a.u., RMS$(|t^{[n]}-t^{[n-1]}|)< 10^{-6}$, and MAX$(|t^{[n]}-t^{[n-1]}|)< 10^{-5}$, where $n$ is the current iteration, RMS represents the root mean square of the difference of the amplitude modulus between successive iterations, and MAX is the maximum value of the difference of the amplitude modulus. The same criteria are used for the $\lambda$ and $\bar t$ amplitudes, where a pseudo-energy is computed as in Eq. \ref{eq:ene}. The convergence of the amplitude equations is accelerated using the direct inversion of the iterative subspace (DIIS) method,\cite{Pulay:DIIS} where the error vector is: $e^{[n]}=t^{[n]}-t^{[n-1]}$, following the work of Scuseria \textit{et al.}.\cite{SCUSERIA1986236} The only difference with the molecular case is that the error matrix is Hermitian rather than symmetric, because $\langle e^{[n]} | e^{[n']}\rangle$ is a complex number. As for the CCSD equations, the formation of the error matrix is transparent for the periodic and molecular cases because \texttt{einsum} can automatically recognize real vs complex arrays. Based on some testing in periodic and molecular calculations, we found that the best compromise between storage requirements and convergence rate is obtained by keeping the amplitudes from five previous iterations and extrapolating every five iterations. The older amplitudes are stored on a binary file and are read in only when performing an extrapolation.

Because NumPy automatically detects BLAS and LAPACK libraries and \texttt{einsum} can optimize the contraction pathway and the reordering of the indices (\texttt{optimization=True} option), our program automatically takes advantage of vectorization and shared memory parallelism for the tensor contractions, which are the most expensive parts of the calculation. The contractions are arranged such that the CCSD $O(N^6)$ scaling is preserved, where here $N=N_fN_k$ is a collective index and $N_f=N_o+N_v$ is the number of basis functions in the simulation cell. As we discussed at the beginning of the section, this is not the most storage/computationally efficient approach but it is the cleanest for this first implementation. One disadvantage of Python is that memory allocation and usage is difficult to control. We implemented a series of checks on the used vs available memory using the \texttt{psutil, tracemalloc} modules, and on Linux platforms we use the \texttt{resource} module to impose a memory limit for the overall program (which is necessary to run, for instance, on a shared cluster where a memory limit is required in the submission script). For now, we only take advantage of this memory information in one instance: the $\tau_{MN}^{AB} \langle MN ||EF \rangle$ contraction for the $\tilde W_{ABEF}$ intermediate in Eq. S9 of the SI. If the available memory is $> 2N_V^4$, then the contraction $\texttt{tau[A,B,M,N]IJAB[E,F,M,N]}$ is performed directly (the indices of the \texttt{tau} and \texttt{IJAB} arrays are reordered explicitly before the contraction, as explained above); otherwise, we set up an external loop over the \texttt{A} index and we perform the contraction on the remaining indices. The advantage of the latter approach is that Python only allocates memory for $N_V^3$ arrays during the contraction, using significantly less memory. This type of approach could be used on $N_ON^3_V$ tensors in the future to further reduce the memory footprint. 

As in Ref. \citenum{hirata_coupled-cluster_2004}, we make the assumption that $\langle PQ|RS \rangle^* = \langle RS|PQ \rangle$ and we permute the indices in the relevant order (as for the $-\langle MA||EF\rangle^*$ and $\langle MN||IE \rangle^* $ integrals in the $\tilde W_{EFAM}$ and $\tilde W_{IEMN}$ intermediates in Eqs. S11-S12 of the SI, respectively); while this assumption is correct in a molecular case with complex orbitals, it is only true in the periodic case if the $k$-point grid is sufficiently fine, which is unlikely for what we can currently afford. At the same time, without this assumption we would have to store explicitly a larger set of 2ERIs in COk basis, further increasing the memory footprint.

\section{Computational Details\label{comp}}

Since we rely on GAUSSIAN to produce the AOr matrices and 2ERI integrals, we do not worry about the number of replica cells in real space explicitly as in Ref. \citenum{hirata_coupled-cluster_2004}. In other words, we let the more advanced algorithms in GAUSSIAN handle the integral screening. The appropriate number of replica cells and $k$ points, integral and CO coefficients accuracy, etc. is first tested on HF calculations in GAUSSIAN to check that the HF energy is converged to the thermodynamic limit. For the system size that we can afford for this periodic CCSD implementation, most of the GAUSSIAN default settings are sufficient: 100 \AA~ for the cell range to evaluate the HF exchange, $10^{-14}$ a.u. accuracy for the 2ERIs, and $10^{-10}$ a.u. accuracy for the HF energy convergence. However, the default number of $k$ points in GAUSSIAN is typically too large for our CCSD code and we select it manually. We use the following notation to indicate the number of $k$ points used: PBC($N$), where $N$ corresponds to the \texttt{NRecip=N} keyword in GAUSSIAN; see the explanation in section \ref{theory} for how the \texttt{N} value is related to the choice of $k$ point mesh. For the polyyne polymer, the $N_C$ value corresponding to the 100 \AA~ cell range ($N_C=41$) was too large for the simulations; thus, we used $N_C=11$ instead, because this value was compatible with the largest $k$-mesh that was affordable for this system: PBC(11).

The PBC calculations are compared against molecular cluster calculations. Since the convergence of the central cell properties with the cluster length is slow, results are reported as differences between two clusters with an odd number of units corresponding to the same central cell as in the PBC calculation, and an equivalent number of units replicated forward and backward. For a quantity $A$, the unit cell value from cluster calculations is obtained as 
\begin{equation}
A=\frac{A_M-A_{M-2}}{2}
\label{eq:cluster}
\end{equation}
where $A_M$ is the quantity obtained in a cluster calculation with $M$ units. Results are also compared with a single-unit (i.e., molecular) calculation. The cluster and molecular CCSD calculations were performed with a development version of GAUSSIAN.\cite{gdv} 

\section{Results\label{results}}

The first system we analyze is a chain of \ce{H2} molecules with a bond distance of 0.74 \AA~ and a translation vector of magnitude $\tilde a =3.0$ \AA~ along the molecular axis (both oriented along the x axis of the coordinate system). The calculations were performed with the 3-21G basis set ($N_o=2$, $N_v=6$). For this system, the number of replica cells in real space is $N_C=37$ and calculations were repeated with a variety of $k$ point meshes.  We consider clusters with 51 and 49 units, and clusters with 37 and 35 units. 

\begin{table*}[htp!]
    \centering
    \begin{tabular}{c|cccc}
        \hline
          & $E(HF)$ & $\Delta E(MP2)$ & $\Delta E(t)$ & $\Delta E(\lambda)$\\
        \hline
        Cluster(51-49) & -1.122231347 & -0.017662931 & -0.02521410 & -0.02483255 \\
        Cluster(37-35) & $< 10^{-9}$ & $< 10^{-9}$ & $-2.50\cdot 10^{-8}$ & $-1.50\cdot 10^{-8}$ \\
        PBC(5)  & $-1.84\cdot 10^{-1}$ & $2.15\cdot 10^{-3}$ & $3.27\cdot 10^{-5}$ & $3.31\cdot 10^{-5}$ \\
        PBC(11) & $-3.53\cdot 10^{-2}$ & $4.61\cdot 10^{-4}$ & $2.06\cdot 10^{-6}$ & $2.05\cdot 10^{-6}$ \\
        PBC(15) & $-2.52\cdot 10^{-2}$ & $3.32\cdot 10^{-4}$ & $8.27\cdot 10^{-7}$ & $8.18\cdot 10^{-7}$ \\
        PBC(21) & $-1.68\cdot 10^{-7}$ & $-1.02\cdot 10^{-7}$ & $-2.17\cdot 10^{-7}$ & $-2.08\cdot 10^{-7}$ \\
        PBC(25) & $<10^{-9}$ & $-4.10\cdot 10^{-9}$ & $-4.07\cdot 10^{-8}$ & $-3.83\cdot 10^{-8}$ \\
        PBC(29) & $< 10^{-9}$ & $-1.80\cdot 10^{-9}$ & $-3.78\cdot 10^{-8}$ & $-3.58\cdot 10^{-8}$ \\
        PBC(33) & $< 10^{-9}$ & $-1.40\cdot 10^{-9}$ & $-3.74\cdot 10^{-8}$ & $-3.54\cdot 10^{-8}$ \\
        Molecular & $-7.09\cdot 10^{-4}$ & $3.50\cdot 10^{-4}$ & $3.41\cdot 10^{-4}$ & $3.22\cdot 10^{-4}$ \\
        \hline
    \end{tabular}
    \caption{Energies (a.u.) for the \ce{H2} chain with the molecular
      axis and translation vector along the x axis of the coordinate
      system ($R_{HH} = 0.74$ \AA, $\tilde a =3.0$ \AA): $E(HF)$ is
      the HF energy, $\Delta E(MP2)$ and $\Delta E(t)$ are the
      correlation energies at MP2 and CCSD levels, respectively, and
      $\Delta E(\lambda)$ is a pseudo-energy for the $\lambda$
      amplitudes computed as in Eq. \ref{eq:ene} with the substitution
      $t\rightarrow \lambda$. The first row reports reference data of
      absolute energies computed as in Eq. \ref{eq:cluster} with
      51-unit and 49-unit clusters. The remaining data are differences
      from the reference. Cluster(37-35) corresponds to a cluster
      calculation involving 37 and 35 units; PBC($N$) represents a PBC
      calculation with \texttt{NRecip=N}; Molecular corresponds to
      data for a single \ce{H2} molecule.}
    \label{tab:H2-ene}
\end{table*}

The HF energy, the 2nd order M\o llet-Plesset (MP2) and CCSD correlation energies, and the $\lambda$ pseudo-energy (computed as in Eq. \ref{eq:ene}) for the \ce{H2} chain are reported in Table \ref{tab:H2-ene}. The table reports the absolute values of a 51-49 cluster calculation as reference, while the rest of the data are reported as differences from this reference (all in atomic units). The absolute energy values are reported in Table S1 of the SI. Passing from a single molecule to a chain, the unit cell energy increases by 0.7 mHartrees, likely because at this geometry the electron repulsion between units is larger than attractive contributions (e.g., electron-nuclear interactions). However, the inclusion of electron correlation reduces this effect by half. Calculations with smaller clusters are in essentially perfect agreement with the reference; as a reminder, the PBC calculations used 37 replica cells for the AOr quantities, including the 2ERIs. On the other hand, PBC calculations agree with the reference only starting from the PBC(21) option, where the difference from the reference is of the order of $10^{-7}$ Hartrees for all methods. A larger number of $k$ points further reduces the difference to numerical noise. For the PBC(5)-PBC(15) calculations, the difference in HF energy is large, of the order of $10^{-1}-10^{-2}$ Hartrees, considerably larger than the difference between single molecule and chain. Somewhat surprisingly, the post-HF quantities show a smaller difference than the HF energy: $10^{-3}-10^{-4}$ Hartrees for MP2 and $10^{-5}-10^{-7}$ Hartrees for CCSD with a small number of $k$ points, and they are basically in perfect agreement with the reference when using finer meshes ($10^{-8}-10^{-9}$ Hartrees).

\begin{table}[htp!]
    \centering
    \begin{tabular}{c|cccc}
        \hline
          & 1000nm & 700nm & 500nm & 300nm\\
        \hline
        Cluster(51-49) & 7.395 & 7.458 & 7.582 & 8.066 \\
        Cluster(37-35) & $<0.001$ & $<0.001$ & $<0.001$ & $<0.001$ \\
        PBC(5)  & -2.386 & -2.407 & -2.449 & -2.610 \\
        PBC(11) & -0.618 & -0.624 & -0.636 & -0.683 \\
        PBC(15) & -0.454 & -0.459 & -0.468 & -0.504 \\
        PBC(21) & -0.027 & -0.028 & -0.030 & -0.037 \\
        PBC(25) & -0.023 & -0.024 & -0.026 & -0.032 \\
        PBC(29) & -0.023 & -0.024 & -0.026 & -0.032 \\
        PBC(33) & -0.023 & -0.024 & -0.026 & -0.032 \\
        Molecular & -1.570 & -1.597 & -1.649 & -1.861 \\
        \hline
    \end{tabular}
    \caption{$\alpha_{xx}$ polarizability (a.u.) for the \ce{H2} chain with the molecular axis and translation vector along the x axis of the coordinate system ($R_{HH} = 0.74$ \AA, $\tilde a =3.0$ \AA) at various wavelengths (nm). The first row reports reference data of $\alpha_{xx}$ computed as in Eq. \ref{eq:cluster} with 51-unit and 49-unit clusters. The remaining data are differences from the reference. Cluster(37-35) corresponds to a cluster calculation involving 37 and 35 units; PBC($N$) represents a PBC calculation with \texttt{NRecip=N}; Molecular corresponds to data for a single \ce{H2} molecule.}
    \label{tab:H2-alpha}
\end{table}

Polarizability calculations were performed at frequencies corresponding to four wavelengths: 1000nm, 700nm, 500nm, and 300nm, and the results are reported in Table \ref{tab:H2-alpha}. The absolute polarizability values  are reported in Table S2 of the SI. For this system, only the tensor element along the molecular axis is nonzero, $\alpha_{xx}$ in the table. Also for this property, the cluster calculations are all in agreement. The chain conformation increases the polarizability by $1.6-1.9$ a.u. compared to the molecular case. For PBC calculations with a small number of $k$ points, PBC(5)-PBC(15), there is a significant difference from the reference; with PBC(5), this difference is even greater than that between the chain and molecular cases, providing a qualitatively wrong trend for the change in polarizability. For PBC calculations with $N\ge 21$, the agreement with the reference is very good, on the order of $0.02-0.04$ a.u. Calculations with even values of \texttt{NRecip} provide very similar results to those in Tables \ref{tab:H2-ene}-\ref{tab:H2-alpha} for the same number of $k$ points, and they are reported in Tables S3-S6 of the SI.

The reason for the remaining discrepancy might be due to finite-size effects, both in the cluster calculations and in the PBC ones (including the number of replica cells in real space and $k$ points in reciprocal space), and to the convergence thresholds. However, this is unlikely given the energy data in Table \ref{tab:H2-ene}. The difference is more likely due to a more subtle effect in the definition of the electric dipole integrals in COk basis for periodic calculations. As discussed in more detail in Refs. \citenum{Kudin:2000bq}, these integrals depend on the gradients of the COk coefficients with respect to $k$, which can be expressed in terms of the original coefficients: $\nabla_k \mathbf C(k) = \mathbf {U}_k \mathbf C(k)$, where $\mathbf U_k$ is a coefficient matrix equivalent to the coupled-perturbed matrix used for the derivative of the molecular orbital coefficients with respect to an external perturbation.\cite{POPLE:1979wr, FRISCH:1990tq} This $\mathbf U_k$ matrix is missing some terms on the diagonal elements that are related to the derivative of the arbitrary phases of the COk coefficients $\mathbf C(k)$, which are difficult to compute and they are usually set to zero (they are referred to as \textit{missing integers}).\cite{Kudin:2000bq, Rerat:2021, Balduf:2022} This does not affect LR electric dipole-electric dipole polarizability calculations at HF and DFT level because they only use the off-diagonal block of the dipole integral matrix (and thus of $\mathbf U_k$), but it does affect the calculation of the unit cell dipole moment (for which a formulation in terms of Berry phase is available)\cite{Kudin:2007} and that of LR polarizabilities that include the magnetic dipole and the electric quadrupole moments.\cite{Rerat:2021, Balduf:2022, Forson:2024} However, the magnitude of this effect is currently unclear compared to other sources of error such as choice of density functional and basis set. In the context of LR-CCSD-PBC, the missing integers in $\mathbf U_k$ are relevant even for the electric dipole-electric dipole polarizability in the length gauge because the full dipole integral matrix is needed, as shown in Eq. \ref{eq:rhs}. The quantitative effect of these missing integers needs to be explored at this level of theory. Nonetheless, we emphasize that this is not directly related to the CCSD equations per se, as a complete formulation of the $\mathbf U_k$ matrix would not change any of the equations reported in section \ref{theory}. For this specific test case, this effect is small compared to the choice of $k$ point mesh size. 

\begin{table*}[htp!]
    \centering
    \begin{tabular}{c|cccc}
        \hline
          & $E(HF)$ & $\Delta E(MP2)$ & $\Delta E(t)$ & $\Delta E(\lambda)$\\
        \hline
        Cluster(51-49) & $-1.12253475$ & $-0.01740180$ & $-0.02496550$ & $-0.02459920$ \\
        PBC(5)  & $ -1.84\cdot 10^{-1} $  &  $ 2.07\cdot 10^{-3} $  &  $ -4.51\cdot 10^{-5} $  &  $ -4.36\cdot 10^{-5} $  \\
        PBC(11) & $ -3.53\cdot 10^{-2} $  &  $ 4.49\cdot 10^{-4} $  &  $ -2.39\cdot 10^{-6} $  &  $ -2.31\cdot 10^{-6} $ \\
        PBC(15) & $ -2.52\cdot 10^{-2} $  &  $ 3.24\cdot 10^{-4} $  &  $ -8.70\cdot 10^{-7} $  &  $ -8.40\cdot 10^{-7} $  \\
        PBC(21) &  $ -1.00\cdot 10^{-8} $  &  $ -1.00\cdot 10^{-8} $  &  $ -1.00\cdot 10^{-8} $  &  $ -1.00\cdot 10^{-8} $ \\
        PBC(25) & $<10^{-9}$  &  $ -1.00\cdot 10^{-8} $  &  $ -1.00\cdot 10^{-8} $  &  $ -1.00\cdot 10^{-8} $   \\
        PBC(29) & $<10^{-9} $  &  $<10^{-9} $  &  $ -1.00\cdot 10^{-8} $  &  $ -1.00\cdot 10^{-8} $   \\
        PBC(33) & $<10^{-9} $  &  $<10^{-9} $  &  $ -1.00\cdot 10^{-8} $  &  $ -1.00\cdot 10^{-8} $  \\
        Molecular & $ -4.06\cdot 10^{-4} $  &  $ 8.87\cdot 10^{-5} $  &  $ 9.26\cdot 10^{-5} $  &  $ 8.84\cdot 10^{-5} $    \\
        \hline
    \end{tabular}
    \caption{Energies (a.u.) for the \ce{H2} chain with the molecular
      axis along the x axis and translation vector along the y axis of
      the coordinate system ($R_{HH} = 0.74$ \AA, $\tilde a =3.0$
      \AA): $E(HF)$ is the HF energy, $\Delta E(MP2)$ and $\Delta
      E(t)$ are the correlation energies at MP2 and CCSD levels,
      respectively, and $\Delta E(\lambda)$ is a pseudo-energy for the
      $\lambda$ amplitudes computed as in Eq. \ref{eq:ene} with the
      substitution $t\rightarrow \lambda$. The first row reports
      reference data of absolute energies computed as in
      Eq. \ref{eq:cluster} with 51-unit and 49-unit clusters. The
      remaining data are differences from the reference. PBC($N$)
      represents a PBC calculation with \texttt{NRecip=N}; Molecular
      corresponds to data for a single \ce{H2} molecule. }
    \label{tab:H2-3y-ene}
\end{table*}

The second system is another chain of \ce{H2} molecules, but this time
with the translation vector (y axis) perpendicular to the molecular
axis (x axis). We will refer to this chain as \ce{H2}-y for
clarity. As for the previous \ce{H2} chain, the calculations were
performed with the 3-21G basis set ($N_o=2$, $N_v=6$) and the number
of replica cells in real space is $N_C=37$. The energy differences
with respect to the 51-49 cluster calculations are reported in Table
\ref{tab:H2-3y-ene}, while the raw energy data are reported in Table
S7 of the SI. Because of the different orientation of the translation
vector, the interaction between molecular units in this case is
slightly smaller than for the previous chain, as shown by the smaller
energy difference between the molecular and the cluster calculations
at every level of theory (compare the last rows in Tables
\ref{tab:H2-ene} and \ref{tab:H2-3y-ene}). Consequently, the
convergence towards the thermodynamic limit of the PBC calculations
with the number of $k$ points is also slightly faster for \ce{H2}-y
compared to the first one (again, compare the data in Tables
\ref{tab:H2-ene} and \ref{tab:H2-3y-ene}). However, the trends between
the two chains are very similar.

\begin{table*}[htp!]
    \centering
    \begin{tabular}{c|cccc|cccc}
        \hline
        & \multicolumn{4}{c|}{$\alpha_{xx}$} & \multicolumn{4}{c}{$\alpha_{yy}$} \\
          & 1000nm & 700nm & 500nm & 300nm & 1000nm & 700nm & 500nm & 300nm \\
        \hline
        Cluster (51-49) & 5.421   & 5.452   & 5.514   & 5.750   & 0.066   & 0.066   & 0.067   & 0.067 \\
        PBC(5) & $ 0.001 $  &  $ 0.001 $  &  $ 0.001 $  &  $<0.001$ &  $ -0.014 $ &  $ -0.014 $ &  $ -0.014 $  &  $ -0.014 $ \\
        PBC(11) & $<0.001$  &  $ 0.001 $  &  $ 0.001 $  &  $<0.001$ &  $ -0.004 $ &  $ -0.004 $ &  $ -0.004 $  &  $ -0.004 $ \\
        PBC(15) & $<0.001$  &  $ 0.001 $  &  $ 0.001 $  &  $<0.001$ &  $ -0.003 $ &  $ -0.003 $ &  $ -0.003 $  &  $ -0.003 $ \\
        PBC(21) & $<0.001$  &  $ 0.001 $  &  $ 0.001 $  &  $<0.001$ &  $ -0.001 $ &  $ -0.001 $ &  $ -0.001 $  &  $ -0.001 $   \\
        PBC(25) & $<0.001$  &  $ 0.001 $  &  $ 0.001 $  &  $<0.001$ &  $ -0.001 $ &  $ -0.001 $ &  $ -0.001 $  &  $ -0.001 $ \\
        PBC(29) & $<0.001$  &  $ 0.001 $  &  $ 0.001 $  &  $<0.001$ &  $ -0.001 $ &  $ -0.001 $ &  $ -0.001 $  &  $ -0.001 $  \\
        PBC(33) & $ 0.001 $  &  $ 0.001 $  &  $ 0.001 $  &  $<0.001$ &  $ -0.001 $ &  $ -0.001 $ &  $ -0.001 $  &  $ -0.001 $ \\
        Molecular & $ 0.404$  &  $ 0.410$  &  $ 0.419 $  &  $ 0.456$ &  $ -0.066$ &  $ -0.66 $ &  $ -0.067 $  &  $ -0.067$ \\
        \hline
    \end{tabular}
    \caption{$\alpha_{xx}$ and $\alpha_{yy}$ polarizability (a.u.) for
      the \ce{H2} chain with the molecular axis along the x axis and
      translation vector along the y axis of the coordinate system
      ($R_{HH} = 0.74$ \AA, $\tilde a =3.0$ \AA) at various
      wavelengths (nm). The first row reports reference data of
      $\alpha_{xx}/\alpha_{yy}$ computed as in Eq. \ref{eq:cluster}
      with 51-unit and 49-unit clusters. The remaining data are
      differences from the reference. PBC($N$) represents a PBC
      calculation with \texttt{NRecip=N}; Molecular corresponds to
      data for a single \ce{H2} molecule.}
    \label{tab:H2-3y-alpha}
\end{table*}

The differences in polarizability for the \ce{H2}-y chain are reported
in Table \ref{tab:H2-3y-alpha} and the raw data are reported in Table
S8 of the SI. There are two important differences between the original
\ce{H2} and \ce{H2}-y chains. First, in \ce{H2}-y, a second element of
the polarizability tensor becomes non-zero, $\alpha_{yy}$; second, the
$\alpha_{xx}$ element in the \ce{H2}-y chain is much closer to the
isolated molecular value than for the original \ce{H2} chain (compare
the last rows in Tables \ref{tab:H2-alpha} and
\ref{tab:H2-3y-alpha}). This indicates that while an interaction
between the molecular units is not negligible for the tensor elements
along the periodic and non-periodic directions, the interaction is
considerably smaller than when the molecular axis is oriented parallel
to the translation vector. As a consequence, the convergence of the
PBC results towards the thermodynamic limit is faster for \ce{H2}-y
than for \ce{H2}. In fact, in the non-periodic direction, the PBC(5)
results are already close to the limit. In the periodic direction,
although 20 $k$ points are necessary to reach convergence, the error
with PBC(11) is already very small (0.004 a.u.), see Table
\ref{tab:H2-3y-alpha}. Notably, the effect of the missing integers
seems rather small for the \ce{H2}-y chain, as the difference between
PBC and reference data is within the range of finite-size effects and
numerical noise from the convergence of the amplitude equations.

\begin{table*}[htp!]
    \centering
    \begin{tabular}{c|cccc}
        \hline
          & $E(HF)$ & $\Delta E(MP2)$ & $\Delta E(t)$ & $\Delta E(\lambda)$\\
        \hline
        Cluster(35-33) & $-7.876549466$ & $-0.012535059$ & $-0.019375530$ & $-0.018952680$ \\
        Cluster(23-21) & $7.27 \cdot 10^{-6 }$ & $-7.36 \cdot 10^{-8 }$ & $-3.00 \cdot 10^{-7 }$ & $-2.80 \cdot 10^{-7}$\\
        PBC(5)  & $-1.59 \cdot 10^{-1 }$ & $1.22 \cdot 10^{-3 }$ & $-1.28 \cdot 10^{-5 }$ & $-9.89 \cdot 10^{-6}$ \\
        PBC(11) & $-4.23 \cdot 10^{-2 }$ & $3.52 \cdot 10^{-4 }$ & $3.66 \cdot 10^{-6 }$ & $3.22 \cdot 10^{-6}$ \\
        PBC(15) & $-5.16 \cdot 10^{-6 }$ & $-4.43 \cdot 10^{-7 }$ & $-8.09 \cdot 10^{-7 }$ & $-6.87 \cdot 10^{-7}$ \\
        PBC(17) & $-5.16 \cdot 10^{-6 }$ & $-2.15 \cdot 10^{-7 }$ & $-3.29 \cdot 10^{-7 }$ & $-2.86 \cdot 10^{-7}$ \\
        PBC(21) & $-5.16 \cdot 10^{-6 }$ & $-1.80 \cdot 10^{-7 }$ & $-2.59 \cdot 10^{-7 }$ & $-2.24 \cdot 10^{-7}$ \\
        Molecular & $1.47 \cdot 10^{-2 }$ & $-3.69 \cdot 10^{-4 }$ & $-1.07 \cdot 10^{-3 }$ & $-9.64 \cdot 10^{-4}$ \\
        \hline
    \end{tabular}
    \caption{Energies (a.u.) for the \ce{LiH} chain with the molecular axis and translation vector along the x axis of the coordinate system ($R_{LiH} = 1.6$ \AA, $\tilde a =5.0$ \AA): $E(HF)$ is the HF energy, $\Delta E(MP2)$ and $\Delta E(t)$ are the correlation energies at MP2 and CCSD levels, respectively, and $\Delta E(\lambda)$ is a pseudo-energy for the $\lambda$ amplitudes computed as in Eq. \ref{eq:ene} with the substitution $t\rightarrow \lambda$. The first row reports reference data of absolute energies computed as in Eq. \ref{eq:cluster} with 35-unit and 33-unit clusters. The remaining data are differences from the reference. Cluster(23-21) corresponds to a cluster calculation involving 23 and 21 units; PBC($N$) represents a PBC calculation with \texttt{NRecip=N}; Molecular corresponds to data for a single \ce{LiH} molecule.}
    \label{tab:LiH-5A-ene}
\end{table*}

The third test case is a \ce{LiH} chain with a bond length of 1.6 \AA~ and a translation vector of $\tilde a = 5.0$ \AA~ along the bond axis (both along the x axis of the coordinate system). The calculations used the STO-3G basis set and all electrons were correlated ($N_o=4$, $N_v=8$). For this geometry, $N_C=23$. The energy data for 35-33 and 23-21 clusters, the molecular case, and PBC chains with various $k$ space meshes are reported in Table \ref{tab:LiH-5A-ene}. The absolute energy values  are reported in Table S9 of the SI. For this system, there is an attractive interaction in the chain that lowers the HF energy by 15 mHartrees compared to the isolated molecule. The correlation energy changes in the opposite direction, but the change is one order of magnitude smaller than the HF energy at CCSD level, and two orders of magnitude smaller at MP2 level. The difference between the two sets of cluster calculations is about $7\cdot 10^{-6}$ Hartrees for the HF energy, but it is one order of magnitude smaller for the CCSD correlation energy (and $\lambda$ pseudo-energy) and two orders of magnitude smaller for the MP2 correlation energy. This trend is similar to that shown by the PBC calculations, from PBC(15) and above. This again indicates that the correlation energy converges faster towards the thermodynamic limit than the HF energy. We emphasize that the PBC HF energy is already converged at PBC(17), as tripling the values of $N_C$ and $N_k$ does not change the result (not reported). Therefore, the difference between cluster and PBC results is due to the former in this case, and much larger clusters should be used to improve the agreement. With a small number of $k$ points, PBC(5) and (11), the agreement of the HF energy with the reference is poor, with differences on the order of $10^{-1}-10^{-2}$ Hartrees, thus similar or larger than the difference between chain and isolated molecule. Remarkably, however, the agreement for the correlation energy is very good, with differences of the order of $10^{-3}-10^{-4}$ Hartrees for MP2 and $10^{-5}-10^{-6}$ Hartrees for CCSD. With finer $k$ space meshes, the PBC correlation energy differs from the reference by about $2-3\cdot 10^{-7}$ Hartrees.

\begin{table*}[htp!]
    \centering
    \begin{tabular}{c|ccc|ccc}
        \hline
        & \multicolumn{3}{c|}{$\alpha_{xx}$} & \multicolumn{3}{c}{$\alpha_{yy/zz}$} \\
          & 1000nm & 700nm & 500nm & 1000nm & 700nm & 500nm \\
        \hline
        Cluster(35-33) & 24.778 & 26.229 & 29.564 & 17.329 & 18.197 & 20.096 \\
        Cluster(23-21) & -0.011 & -0.012 & -0.013 & 0.002 & 0.002 & 0.003 \\
        PBC(5)  & -12.840 &-13.575 & -15.230 & 0.022 & 0.025 & 0.032 \\
        PBC(11) & -4.818 & -5.145 & -5.892 & 0.000 & 0.001 & 0.002 \\
        PBC(15) & -2.966 & -3.180 & -3.664 & -0.003 & -0.002 & -0.002 \\
        PBC(17) & -2.967 & -3.181 & -3.667 & -0.004 & -0.004 & -0.004 \\
        PBC(21) & -2.968 & -3.182 & -3.668 & -0.005 & -0.005 & -0.005 \\
        Molecular & -13.196 & -13.144 & -11.784 & 6.018 & 6.818 & 8.811 \\
        \hline
    \end{tabular}
    \caption{$\boldsymbol{\alpha}$ polarizability (a.u.) for the
      \ce{LiH} chain with the molecular axis and translation vector
      along the x axis of the coordinate system ($R_{LiH} = 1.6$ \AA,
      $\tilde a =5.0$ \AA) at various wavelengths (nm). The first row
      reports reference data of $\alpha$ computed as in
      Eq. \ref{eq:cluster} with 35-unit and 33-unit clusters. The
      remaining data are differences from the
      reference. Cluster(23-21) corresponds to a cluster calculation
      involving 23 and 21 units; PBC($N$) represents a PBC calculation
      with \texttt{NRecip=N}; Molecular corresponds to data for a
      single \ce{LiH} molecule.}
    \label{tab:LiH-5-alpha}
\end{table*}

The results for the polarizability calculations for the first \ce{LiH} chain are collected in Table \ref{tab:LiH-5-alpha}. The absolute polarizability values  are reported in Table S10 of the SI. There are two distinguished values of polarizability: $\alpha_{xx}$ and $\alpha_{yy}=\alpha_{zz}$. We do not consider the wavelength at 300nm because it is in the resonance region for the molecular case and in the pre-resonance region for the chain. The cluster results indicate that the polarizability along the molecular axis is larger than in the orthogonal directions: $\alpha_{xx}> \alpha_{yy}$ (the two sets of cluster calculations are in good agreement with each other). This is a qualitatively opposite compared to the molecular case, where $\alpha_{xx} < \alpha_{yy}$. The increase in polarizability along the periodic direction (same as the molecular axis) from isolated molecule to chain is twice as large as the decrease in polarizability in the orthogonal directions. For the PBC calculations, there is very good agreement with the reference for the polarizability elements in the non-periodic directions, even with very small $k$ space meshes. This difference is of the same magnitude as the difference between the two sets of cluster calculations, indicating that this is a finite-size effect likely on the part of the cluster calculations, similar to the energy case discussed in the previous paragraph. There are more substantial differences between the cluster and PBC results for the $\alpha_{xx}$ element. Even with the larger $k$ meshes, there is difference of $3-3.5$ a.u. for all wavelengths. This effect is most likely due to the missing integers issue discussed above. Nonetheless, the magnitude of the effect is small compared to the changes from molecule to chain ($12-13$ a.u.) and to the effect of $k$ point mesh. The latter is particularly significant with PBC(5), with differences from the reference around $13-15$ a.u.

\begin{table*}[htp!]
    \centering
    \begin{tabular}{c|cccc}
        \hline
          & $E(HF)$ & $\Delta E(MP2)$ & $\Delta E(t)$ & $\Delta E(\lambda)$\\
        \hline
        Cluster(35-33) & -7.925283198 & -0.013225332 & -0.018918630 & -0.018571415 \\
        PBC(5)  & $-3.36 \cdot 10^{-1}$ & $2.20 \cdot 10^{-3}$ & $-5.06 \cdot 10^{-5}$ & $-4.35 \cdot 10^{-5}$ \\
        PBC(11) & $-6.45 \cdot 10^{-2}$ & $4.86 \cdot 10^{-4}$ & $-1.45 \cdot 10^{-5}$ & $-1.29 \cdot 10^{-5}$ \\
        PBC(15) & $-4.61\cdot 10^{-2}$ & $3.54\cdot 10^{-4}$ & $-4.93\cdot 10^{-6}$ & $-4.49\cdot 10^{-6}$  \\ 
        PBC(17) & $-4.03 \cdot 10^{-2}$ & $3.30 \cdot 10^{-4}$ & $9.21 \cdot 10^{-6}$ & $8.98 \cdot 10^{-6}$ \\
        PBC(21) & $-3.63 \cdot 10^{-5}$ & $-1.02 \cdot 10^{-5}$ & $-1.25 \cdot 10^{-5}$  & $-1.12 \cdot 10^{-5}$ \\
        PBC(25) & $-3.62 \cdot 10^{-5}$ & $-5.92 \cdot 10^{-6}$ & $-5.98 \cdot 10^{-6}$ & $-5.41 \cdot 10^{-6}$ \\
        Molecular & $6.51 \cdot 10^{-2}$ & $-2.97 \cdot 10^{-5}$ & $-2.26 \cdot 10^{-3}$ & $-2.02 \cdot 10^{-3}$ \\
        \hline
    \end{tabular}
    \caption{Energies (a.u.) for the \ce{LiH} chain at the optimized geometry with the molecular axis and translation vector along the x axis of the coordinate system ($R_{LiH} = 1.642252$ \AA, $\tilde a =3.284483$ \AA): $E(HF)$ is the HF energy, $\Delta E(MP2)$ and $\Delta E(t)$ are the correlation energies at MP2 and CCSD levels, respectively, and $\Delta E(\lambda)$ is a pseudo-energy for the $\lambda$ amplitudes computed as in Eq. \ref{eq:ene} with the substitution $t\rightarrow \lambda$. The first row reports reference data of absolute energies computed as in Eq. \ref{eq:cluster} with 35-unit and 33-unit clusters. The remaining data are differences from the reference. PBC($N$) represents a PBC calculation with \texttt{NRecip=N}; Molecular corresponds to data for a single \ce{LiH} molecule.}
    \label{tab:LiH-opt-ene}
\end{table*}

To address how geometrical parameters may affect this property, we optimized the LiH chain geometry at HF-PBC/STO-3G level. The optimized chain has a bond length of 1.642252 \AA~ and a translation vector along the molecular axis (x axis) of magnitude $\tilde a =3.284483$ \AA. Therefore, while the bond length increased slightly by 0.042 \AA, the intermolecular distance decreased substantially by 1.716 \AA. At this geometry, $N_C=35$. The energy data at this new geometry are reported in Table \ref{tab:LiH-opt-ene}. The absolute energy values  are reported in Table S11 of the SI. The HF energy for the 35-33 cluster decreased by almost 50 mHartrees while the corresponding molecular energies differ by only 1 mHartree (we do not report 23-21 cluster calculations in this case as the cluster size is too small). The change in correlation energy is much smaller than that of the HF energy as the geometry changes: it is lower by $<1$ mHartree at the optimized geometry at MP2 level and it is higher by about the same amount at CCSD level. This behavior indicates that electron correlation is mostly a local effect on the molecular unit rather than an intermolecular effect, at least with such a small basis set. In terms of PBC calculations, a larger number of $k$ points is needed to achieve a good agreement with the reference for the HF energy, i.e., PBC(21) and above. The correlation energy is less sensitive, with a difference of $10^{-4}-10^{-5}$ Hartrees already with PBC(17). However, the agreement with the reference is never $<10^{-6}$ Hartrees, probably due to finite-size effects on both the molecular cluster and PBC calculations.

\begin{table*}[htp!]
    \centering
    \begin{tabular}{c|ccc|ccc}
        \hline
        & \multicolumn{3}{c|}{$\alpha_{xx}$} & \multicolumn{3}{c}{$\alpha_{yy/zz}$} \\
          & 1000nm & 700nm & 500nm & 1000nm & 700nm & 500nm \\
        \hline
        Cluster(35-33) & 20.059 & 20.503 & 21.397 & 12.403 & 12.746 & 13.448 \\
        PBC(5)  & -7.450 & -7.704 & -8.215 & 0.102 & 0.110 & 0.127 \\
        PBC(11) & -4.116 & -4.253 & -4.529 & 0.009 & 0.010 & 0.012 \\
        PBC(15) & -3.808 & -3.933 & -4.186 & 0.002 & 0.003 & 0.003 \\
        PBC(17) & -3.528 & -3.643 & -3.875 & 0.015 & 0.015 & 0.017 \\
        PBC(21) & -2.678 & -2.776 & -2.974 & 0.004 & 0.005 & 0.006 \\
        PBC(25) & -2.703 & -2.801 & -3.000 & 0.001 & 0.002 & 0.002 \\
        Molecular & -7.220 & -5.915 & -1.191 & 11.510 & 12.935 & 16.392 \\
        \hline
    \end{tabular}
    \caption{$\boldsymbol{\alpha}$ polarizability (a.u.) for the \ce{LiH} chain with the molecular axis and translation vector along the x axis of the coordinate system ($R_{LiH} = 1.642252$ \AA, $\tilde a =3.284483$ \AA) at various wavelengths (nm). The first row reports reference data of $\alpha$ computed as in Eq. \ref{eq:cluster} with 35-unit and 33-unit clusters. The remaining data are differences from the reference. PBC($N$) represents a PBC calculation with \texttt{NRecip=N}; Molecular corresponds to data for a single \ce{LiH} molecule.}
    \label{tab:LiH-opt-alpha}
\end{table*}

The polarizability data are reported in Table \ref{tab:LiH-opt-alpha}. The absolute polarizability values  are reported in Table S12 of the SI. The reference values are smaller in magnitude by $5-8$ a.u. with respect to the previous geometry, while the molecular values changed by $<1$ a.u. Therefore, there is a smaller difference between molecule and chain for the $\alpha_{xx}$ element at the optimized geometry compared to the previous case, but a larger difference for the $\alpha_{yy}/\alpha_{zz}$ elements (compare Tables \ref{tab:LiH-5-alpha} and \ref{tab:LiH-opt-alpha}). For the PBC calculations, the agreement with the reference is very good along the non-periodic directions, but a larger number of $k$ points are required to get to convergence compared to the previous geometry. For the periodic direction, the difference between the PBC data and the reference is of the order of $2.7-3.0$ a.u. for PBC(25), similar in magnitude to the $\tilde a =5.0$ \AA~ configuration. These results support the hypothesis that this difference is due to the missing integers issue, as the trends with changes in geometry are correct for periodic and non-periodic directions.

The last system we consider is a polyyne polymer, with the molecular
axis along the x axis of the coordinate system ($R_{CC} = 1.2012$ \AA,
$\tilde a =2.7412$ \AA). The calculations were performed with the
STO-3G basis set ($N_o=12$, $N_v=6$) and $N_C=11$. Note that $N_C$ is
smaller than the default value for this system in GAUSSIAN ($N_C=41$),
but we are limited by the efficiency of CCResPy for this system. For
the same reason, we limited the PBC calculations to \texttt{NRecip=9}
and \texttt{NRecip=11}. The cluster calculations include sets of 21-19
units and 11-9 units, in both cases with H atoms capping the ends of
the finite clusters ($R_{CH} = 1.07$ \AA). Therefore, in this system
we expect larger finite-size errors.

\begin{table*}[htp!]
    \centering
    \begin{tabular}{c|cccc}
        \hline
          & $E(HF)$ & $\Delta E(MP2)$ & $\Delta E(t)$ & $\Delta E(\lambda)$\\
        \hline
        Cluster (21-19) & $-74.72469349$ & $-0.14602053$ & $-0.15618130$ & $-0.15164120$ \\
        Cluster (11-9) & $ -1.36\cdot 10^{-5} $  &  $ 1.06\cdot 10^{-5} $  &  $ 1.18\cdot 10^{-5} $  &  $ 1.10\cdot 10^{-5} $ \\
        PBC(9) & $ 3.07\cdot 10^{-6} $  &  $ -3.51\cdot 10^{-4} $  &  $ -2.59\cdot 10^{-4} $  &  $ -2.30\cdot 10^{-4} $   \\
        PBC(11) & $ 2.58\cdot 10^{-6} $  &  $ -2.44\cdot 10^{-4} $  &  $ -1.63\cdot 10^{-4} $  &  $ -1.55\cdot 10^{-4} $  \\
        Molecular & $ -1.13\cdot 10^{-0} $  &  $ -4.62\cdot 10^{-3} $  &  $ -1.35\cdot 10^{-2} $  &  $ -1.43\cdot 10^{-2} $   \\
        \hline
    \end{tabular}
    \caption{Energies (a.u.) for the polyyne polymer  ($R_{CC} = 1.2012$ \AA, $\tilde a =2.7412$ \AA): $E(HF)$ is the HF energy, $\Delta E(MP2)$ and $\Delta E(t)$ are the correlation energies at MP2 and CCSD levels, respectively, and $\Delta E(\lambda)$ is a pseudo-energy for the $\lambda$ amplitudes computed as in Eq. \ref{eq:ene} with the substitution $t\rightarrow \lambda$. The first row reports reference data of absolute energies computed as in Eq. \ref{eq:cluster} with 21-unit and 19-unit clusters. The remaining data are differences from the reference. Cluster(11-9) corresponds to a cluster calculation involving 11 and 9 units; PBC($N$) represents a PBC calculation with \texttt{NRecip=N}; Molecular corresponds to data for a single \ce{C2H2} molecule.}
    \label{tab:poly-ene}
\end{table*}

The energy difference data are reported in Table \ref{tab:poly-ene}
and the raw data are reported in Table S13 of the SI. The difference
between the 21-19 and 11-9 cluster is of the order of $10^{-5}$
Hartrees for the HF energy and the correlation energy. On the other
hand, the difference between a single molecular units and the clusters
is substantial: about 1 Hartree for the HF energy and $10^{-2}$
Hartrees for the correlation energy. This suggests that while the
finite-size error is larger than for the chain systems studied up to
this point, it is not dramatic. The PBC energy differences are of the
order of $10^{-6}$ Hartrees for the HF energy, but they are larger for
correlation energy: $10^{-4}$ Hartrees. This indicates that electron
correlation is stronger for this conjugated polymer than for the
molecular chains and a finer $k$ mesh would be necessary to reach the
thermodynamic limit.

\begin{table*}[htp!]
    \centering
    \begin{tabular}{c|ccc|ccc}
        \hline
        & \multicolumn{3}{c|}{$\alpha_{xx}$} & \multicolumn{3}{c}{$\alpha_{yy/zz}$} \\
          & 1000nm & 700nm & 600nm  & 1000nm & 700nm & 600nm  \\
        \hline
        Cluster (21-19) & 40.861  & 41.479  & 41.929    & 1.620  & 1.627  & 1.632   \\
        Cluster (11-9) & $ -0.262 $  &  $ -0.273 $  &  $ -0.281 $  &  $<0.001$ &  $<0.001$ &  $<0.001$ \\
        PBC(9) & $ -1.441 $  &  $ -1.448 $  &  $ -1.452 $  &  $ 0.002 $ &  $ 0.002 $ &  $ 0.002 $  \\
        PBC(11) &  $ -1.625 $  &  $ -1.653 $  &  $ -1.670 $  &  $ 0.001 $ &  $ 0.001 $ &  $ 0.001 $  \\
        Molecular & $ -25.231 $  &  $ -25.793 $  &  $ -26.203 $  &  $ 0.359 $ &  $ 0.359 $ &  $ 0.360 $  \\
        \hline
    \end{tabular}
    \caption{$\boldsymbol{\alpha}$ polarizability (a.u.) for the
      polyyne chain ($R_{CC} = 1.2012$ \AA, $\tilde a =2.7412$ \AA) at
      various wavelengths (nm). The first row reports reference data
      of $\boldsymbol{\alpha}$ computed as in Eq. \ref{eq:cluster}
      with 21-unit and 19-unit clusters. The remaining data are
      differences from the reference. Cluster(11-9) corresponds to a
      cluster calculation involving 11 and 9 units; PBC($N$)
      represents a PBC calculation with \texttt{NRecip=N}; Molecular
      corresponds to data for a single \ce{C2H2} molecule.}
    \label{tab:poly-alpha}
\end{table*}

The data of polarizability differences are reported in Table
\ref{tab:poly-alpha} for three wavelengths (1000nm, 700nm, and 600nm),
while the raw data are in Table S14 of the SI. Because of the symmetry
of this polymer, the tensor elements in the non-periodic directions
are equal to each other and much smaller than that in the periodic
direction: $\alpha_{xx}>\alpha_{yy}=\alpha_{zz}$. This trend is
similar for the single molecular unit, but the magnitude is less than
half of that of the polymer for the large polarizability element. The
finite-size error is likely small, as the difference between the 21-19
and 11-9 cluster results is $<1\%$ for $\alpha_{xx}$ and it is
negligible for $\alpha_{yy/zz}$. Assuming a similar order of magnitude
for the finite-size effect of the PBC calculations (i.e., assuming
that the 11-9 cluster and PBC(11) calculations are roughly
equivalent), most of the error along the periodic direction is due to
the missing integer issue as for the LiH case. However, this effect
for the polyyne polymer is much smaller in magnitude than for the LiH
chain, as the error with respect to the reference is 4\% for polyyne
and 14\% for LiH. On the other hand, the error along the non-periodic
directions is again negligible and due to the finite-size effect on
both the cluster and PBC calculations.

\section{Discussion and Conclusions\label{discuss}}

In this work, we present the first implementation of the frequency-dependent electric dipole-electric dipole polarizability for 1D periodic systems at CCSD level in the CCResPy open source code.\cite{github-ccrespy} The implementation is based on the Python NumPy module library. This paper presents complete equations  and many details of the implementation. One of the code's main features is that the integrals and amplitudes use a collective-index notation, shown in Table \ref{tab:index}. This choice allows us to write the equations and the corresponding code in a transparent way for molecular and periodic cases. The advantage is that the code is cleaner, easier to read, and further development can be carried out simultaneously for molecules and periodic system (even possibly for a molecular case with complex orbitals). The drawback is that it does not take advantage of the momentum conservation conditions in Eq. \ref{eq:momentum} for storage and computational cost savings. Similar considerations apply for the choice of treating all spin cases together. Nonetheless, CCResPy uses efficient tensor contraction libraries, vectorization, and shared memory parallelism implicit in the NumPy package. Furthermore, we implemented memory limits and checks to run safely on super-computing clusters with a queuing system.

There are many improvements that can be incorporated after this initial implementation. A few obvious examples include a restricted closed-shell implementation and an explicit summation over $k$ points that takes advantage of momentum conservation to avoid summations over forbidden indices. Other possible improvements include saving more large arrays on disk, e.g., $N_ON_V^3$ integrals and intermediates, rather than keeping them in memory, and implementing I/O operations in batches to reduce memory requirements. More complicated improvements may include performing contractions in mixed AOr/COk bases, akin to integral-direct contractions in molecular codes. 

In terms of features, some planned developments include the extension of periodicity in 3D, and the implementation of mixed polarizabilities that include magnetic dipoles and higher multipole perturbations.\cite{Balduf:2022} The LR code may also be extended to compute excitation energies directly as poles of the LR function. Further out developments may include the implementation of higher-order response functions. 

The simulations presented in this study include some proof-of-concept examples that showcase the capabilities of the code, and provide physical insight on the effect of intermolecular interactions on an important optical property of molecular materials, i.e., the frequency-dependent electric dipole-electric dipole polarizability. These calculations also offer critical insight on the performance of periodic LR-CCSD with respect to the reciprocal space sampling. The calculations in section \ref{results} show that the correlation energy is less sensitive than the HF energy to the number of $k$ points for the molecular chains (\ce{H2}, \ce{H2}-y, and LiH): sub-mHartree accuracy in the correlation energy can be achieved already with 10 $k$ points (including edge and $\Gamma$ points of the first Brillouin zone). This is an indication that electron correlation is mostly a local effect for these systems, which is reasonable considering that these are chains made out of isolated molecular units and the basis sets used for the simulations are small and compact. For the conjugated polymer polyyne, the effect of electron correlation is larger than for the molecular chains, and the agreement with the cluster calculations is two orders of magnitude better for the HF energy than for the correlation energy, see Table \ref{tab:poly-ene}. Nonetheless, even with PBC(9) and PBC(11), the difference of the correlation energy from the reference is below 1 mHartree. However, the polarizability requires a larger number of $k$ points to obtain reasonable accuracy (here somewhat loosely defined as differences of the order of $10^{-2}$ a.u. in polarizability when increasing the number of $k$ points). More specifically, while PBC(5) CCSD correlation energy values are within $10^{-5}$ Hartrees from the cluster reference data for all chain systems, see Tables \ref{tab:H2-ene}, \ref{tab:H2-3y-ene}, \ref{tab:LiH-5A-ene}, and \ref{tab:LiH-opt-ene}, the polarizability shows qualitatively incorrect trends compared to the molecule $\rightarrow$ chain change, see Tables \ref{tab:H2-alpha}, \ref{tab:H2-3y-alpha}, \ref{tab:LiH-5-alpha}, and \ref{tab:LiH-opt-alpha}. PBC(11) polarizability data are qualitatively correct, but the difference from the reference data is ~2-10 times larger than that obtained with converged $k$-point calculations. For polyyne, we do not have a sufficient number of $k$-mesh options to draw definitive conclusions about convergence towards the thermodynamic limit, but considering the 11-9 cluster and PBC(11) calculations roughly equivalent in quality, the finite-size error should be fairly small for this system even for the tensor element along the periodic axis ($\alpha_{xx}$), see Table \ref{tab:poly-alpha}.

While the calculations of polarizability values in non-periodic directions are in very good agreement with the cluster reference values (within finite-size constraints), the calculations along the periodic directions provide larger differences, at least for the \ce{LiH} chains and the polyyne polymer. We attribute this difference to a limitation in the definition of the periodic electric dipole moment in the LG, discussed in section \ref{results} and in Refs. \citenum{Kudin:2000bq,Balduf:2022, Rerat:2021}. The evaluation of these missing integers is non-trivial, and there are some claims that their effect should be alleviated with large basis sets and very fine $k$ space meshes;\cite{desmarais2023first} both things are not currently possible at CCSD level. A different way to sidestep the issue is to compute the polarizability tensor using the velocity gauge (VG), where the dipole operator is represented by the momentum rather than the position operator.\cite{Pedersen:2004eg} In the VG, the $\mathbf{U}_k$ matrix is not required and the missing integers issue disappears. However, the VG and LG formulations are numerically only equivalent for exact methods at the complete basis set limits. We are working on extending CCResPy to compute $\boldsymbol{\alpha}(VG)$ and we will present a LG/VG comparison in future work. Nevertheless, we emphasize that this issue is quantitatively small for the \ce{H2} and \ce{H2}-y cases and even for the polyyne polymer (error about 4\%); the error is reasonably small even for the \ce{LiH} chain (14\%), at least compared with the choice of $k$ mesh. It is also likely that the choice of basis set plays as large a role as that of the $k$ mesh size in terms of accuracy for these simulations. 

A direct comparison of computational cost between the cluster and PBC calculations is not possible, because the former were performed with an optimized standard quantum chemistry code (GAUSSIAN) while the latter were performed with CCResPy. A fairer comparison would have been to also perform the cluster calculations with CCResPy. Although possible in principle, these cluster calculations were too computationally demanding for the current version of our code. Additionally, even the GAUSSIAN cluster calculations on the \ce{LiH} chain became very intensive for sizes beyond those presented in this work and we did not pursue them further. Consider also that the cluster results require separate calculations on two systems that differ by two units (see Eq. \ref{eq:cluster}). Therefore, a qualitative rule-of-thumb is that the LR-CCSD-PBC calculations are considerably faster than the cluster ones for $N_k\le N_C$. The advantage would be even more pronounced for 3D systems.

In summary, we have presented the first implementation of frequency-dependent response properties at 1D-periodic CCSD level, together with an open-source code that can be used and improved by the community. This work may open the door to accurate simulations of optical properties of materials with an approach that is systematically improvable and as reliable as that for molecules.

\section*{Acknowledgements}
M.C., T.P., and J.A. gratefully acknowledge support from the National Science Foundation through Grant No. CHE-2154452. The calculations were
performed at the University of Kansas Center for Research Computing
(CRC), including the BigJay cluster resource funded through NSF Grant
MRI-2117449.

\section*{Supporting Information}
The supporting information contains the equations for all
intermediates amplitudes used in Eqs. \ref{eq:t1}-\ref{eq:lrf} of the
main text. It also contains the absolute values of the energy and of
the polarizability tensor elements for all of the molecular and
periodic calculations.

\bibliography{ccrespy}

\end{document}